\documentclass[a4paper,10pt,prl,twocolumn]{revtex4}
\usepackage[dvips]{graphics,color}
\usepackage{amsmath}
\usepackage{amsfonts}
\newcommand{\one}{\mathrm{I} \! \! 1}
\usepackage{amssymb}
\usepackage{amstext}
\usepackage[sort&compress]{natbib}
\begin{document}

\title{Bounds on Multipartite Entangled Orthogonal State Discrimination Using Local Operations and Classical Communication}

\author{M. Hayashi $^1$}
\author{D. Markham $^2$} \thanks{Authors are listed alphabetically. Corresponding authors are Markham (markham@phys.s.u-tokyo.ac.jp), Virmani
(s.virmani@imperial.ac.uk) and Owari
(owari@eve.phys.s.u-tokyo.ac.jp).}
\author{M. Murao $^{2,3}$}
\author{M. Owari $^{2 *}$}
\author{S. Virmani $^{4 *}$}
\affiliation{$^1$Imai Quantum Computation and Information Project,
ERATO, JST, Tokyo 113-0033, Japan\\
\& Superrobust Computation Project (21st Century COE by MEXT),
University of Tokyo, Japan}

\affiliation{$^2$Department of Physics, Graduate School of
Science,
  University of Tokyo, Tokyo 113-0033, Japan}

\affiliation{$^3$PRESTO, JST, Kawaguchi, Saitama 332-0012, Japan}

\affiliation{$^4$ Optics Section, Blackett Laboratory \& Institute
for Mathematical Sciences, Imperial College, London SW7 2AZ,
United Kingdom}

%
%

%
%

\begin{abstract}
We show that entanglement guarantees difficulty in the
discrimination of orthogonal multipartite states locally. The
number of pure states that can be discriminated by local
operations and classical communication is bounded by the total
dimension over the average entanglement. A similar, general
condition is also shown for pure and mixed states. These results
offer a rare operational interpretation for three abstractly
defined distance like measures of multipartite entanglement.
\end{abstract}
%
\maketitle
The problem of defining and understanding multiparty entanglement
is a major open question in the field of quantum information. As
entanglement theory becomes more useful in other areas of many
body physics, multiparty entanglement becomes increasingly
relevant to general physics, too. Hence, understanding the {\it
meaning} of entanglement has become and interesting and important
question.

In the bipartite case, entanglement is fairly well understood
\cite{Horodecki01}. There are many entanglement measures defined
both operationally (in terms of the usefulness of states for
quantum information tasks) and abstractly (such that they obey
certain axioms and may be called entanglement monotones). One of
the most celebrated results in bipartite entanglement theory is
that for pure states essentially all measures coincide and have
clear operational relevance. For more than two parties however,
the operational approach quickly becomes very difficult. There are
no clear ``units of usefulness'' and we have the possibility of
inequivalent types of entanglement \cite{Dur00}. Some abstract
measures do persist by their simplicity. In particular those
measures that define ``proximity'' to the set of separable states
\cite{Vidal99,Vedral98,Wei03} have natural multiparty analogues.
However, due to their abstract definition, their operational
meaning is not clear and remains an open question.

\begin{figure}[t]
\resizebox{!}{4.3cm}{\includegraphics{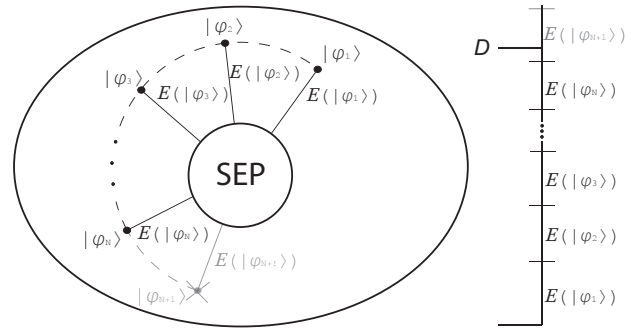}}
\caption{\label{Fig. 1}To discriminate the pure states
$\{|\varphi_i\rangle\}_{i=1}^N$ perfectly under LOCC, the sum of
the entanglement ``distances'' $E(|\varphi_i\rangle)$ must be less
than the total dimension $D$ (Theorem 1 and 2), thus $N\leq
D/\overline{E(|\varphi_i\rangle)}$.}
\end{figure}

In this Letter, we consider the connection between distance-like
entanglement measures and the task of local operations and
classical communication (LOCC) state discrimination with this
question in mind. This task illustrates the restriction of only
having local access to a system, fundamental to the use of
entanglement in quantum information (and notions of locality).
Indeed, LOCC measurement of quantum states is important for
cryptographic protocols \cite{{Bennett84Ekert91}}, channel
capacities \cite{Hayden04Watrous05}, and distributed quantum
information processing \cite{Cirac99}.

Intuitively we expect that entangled states are more difficult to
discriminate locally, since inherently they possess properties
that are non-local. Indeed it is known that entanglement can make
LOCC discrimination more difficult \cite{Terhal01}. But the exact
relation is thus far unclear, and there are no general
quantitative results. The results that are known can be confusing.
One of the earliest results on the subject reveals a set of
non-entangled, product states that cannot be discriminated
perfectly by LOCC \cite{Bennett99}. Later it was shown that any
two pure states can be discriminated optimally by LOCC, no matter
how entangled they are \cite{Walgate00Virmani01}. There have been
several results since then on various LOCC settings
\cite{Chen02-Hillery02}, and connections have been made to
bipartite entanglement distillation and formation
\cite{Badziag03etal}. However many results are specific to the
bipartite case, or only valid for specific scenarios.

We show a clear connection between distancelike measures of
entanglement and LOCC state discrimination in the general
multipartite case. We first show how the conditions imposed on the
measurement by perfect state discrimination can be rewritten in
terms of a quantity which looks like a ``distance'' to the closest
separable state. As we weaken these conditions, we then show that
this relates directly to three entanglement measures. Finally,
combining these results gives a general (pure and mixed state)
bound, and, for pure states, allows the following interpretation:
entanglement gives an upper bound to the number of pure states
that can be discriminated perfectly by LOCC.

By using known entanglement results we will give examples of
existing and new LOCC discrimination bounds in a unified manner.

{\bf Theorem 1:} A necessary condition for deterministic LOCC
discrimination of set $\{\rho_i|i=1..N\}$ is that the following
inequality holds:
\begin{equation}
\sum_i d(\rho_i) \leq D, \label{eqn: Theorem 1}
\end{equation}
where $D$ is the total dimension of the system, and
\begin{eqnarray}
&d(\rho_i) := \min \frac{1}{{\rm tr}\{\rho_i\omega_i\} }\nonumber \\
&{\mbox{such that}} \nonumber \\
&(i)~ \one \geq \frac{\omega_i}{{\rm tr}\{\rho_i\omega_i\}} \geq 0,
~~(ii)~\omega_i \in SEP, \label{eqn: DEF d1}
\end{eqnarray}
where $SEP$ denotes the set of separable operators.

To prove theorem 1, we begin by listing some conditions that the
POVMs (positive operator value measures) must satisfy.
The task of state discrimination is to perform a measurement (in our
case by LOCC) on a system to find out which one of a set of states
the system is in. If it is possible to perfectly discriminate among
a set of density matrices $\mathcal{S}:=\{\rho_i|i=1..N\}$ by LOCC,
then it is necessary that there exists a POVM $\{M_i\}$ satisfying
the following conditions:
\begin{eqnarray}
\sum_i M_i &=& \one \label{eqn: condition completeness}\\
\one \geq M_i &\geq& 0 ~~~~~ \label{eqn: condition positivity} \\
\forall i~~~ {\rm tr}\{M_i \rho_i\} &=& 1 ~~~~~ \label{eqn: condition perfect} \\
\forall i~~~M_i &\in& SEP ~~~~~ \label{eqn: condition separable}
\end{eqnarray}
Conditions (\ref{eqn: condition completeness}) and (\ref{eqn:
condition positivity}) are simply the conditions mean that
$\{M_i\}$ is a POVM. Condition (\ref{eqn: condition perfect}) says
that, given a state $\rho_i$, the result corresponding to outcome
$M_i$ occurs with probability $1$, i.e. the discrimination is
deterministic. Condition (\ref{eqn: condition separable}) is known
to be a necessary condition if the POVM $\{M_i\}$ is to be
implementable by LOCC \cite{Terhal01}.

To make the connection to distances between states we first notice
that any POVM element $M_i$ can be expressed as a positive number
$s_i={\rm tr}\{M_i\}$ times a density matrix $\omega_i$, $M_i =
s_i \omega_i$. We can then use this to immediately rewrite
(\ref{eqn: condition completeness})-(\ref{eqn: condition
separable}). Condition (\ref{eqn: condition perfect}) is rewritten
$s_i=1/{\rm tr}\{\rho_i\omega_i\}$. Condition (\ref{eqn: condition
separable}) means $\omega_i$ is separable. For pure states $s_i$
now looks like a distancelike quantity between state $\rho_i$ and
some separable state $\omega_i$, such that the remaining
conditions are satisfied (that is $\sum_i s_i \omega_i = \one$,
 $\one \geq s_i \omega_i \geq 0$).

If we then minimise $s_i$ such that conditions (\ref{eqn: condition
positivity}), (\ref{eqn: condition perfect}) and (\ref{eqn:
condition separable}) are satisfied for each $i$ independently, we
get exactly the definition of $d(\rho_i)$ in theorem 1 (\ref{eqn:
DEF d1}). Condition (\ref{eqn: condition completeness}) implies this
minimisation must satisfy
\begin{equation}
\sum_i d(\rho_i) \leq D \label{eqn: trace D},
\end{equation}
completing the proof.$\Box$

At this point $d(\rho)$ cannot be considered a `distance to the
closest separable state' entanglement measure. It turns out that
condition (i) in (\ref{eqn: DEF d1}) complicates things a lot, and
indeed, even without this condition it is not an entanglement
monotone for mixed states [see comment below the definition of the
geometric measure (\ref{geometric entanglement})]. Hence the
connection to entanglement is not immediate. However, we can use
this quantity to relate the problem of state discrimination to
other distance-like entanglement monotones, as in the following
theorem.

{\bf Theorem 2:} The following bounds hold for all states $\rho$:
\begin{equation}
d(\rho) \geq r(\rho) \geq 2^{E_R(\rho) + S(\rho)} \geq
2^{G(\rho)}, \label{eqn: Theorem 2}
\end{equation}
where $G(\rho)$ is the geometric measure; $E_R(\rho)$ is the
relative entropy of entanglement; $S(\rho)$ is the von Neumann
entropy; and $r(\rho) := |P|(1+R_G(P/|P|)$, where $P$ is the support
of state $\rho$ \cite{NoteSupport}, $|P|:=$tr$\{P\}$, and
$R_G(\rho)$ is the robustness of entanglement of state $\rho$.

In the pure state case $S(\rho)=0$ and $P = \rho$, and so these
quantities become exactly (up to log) the geometric measure of
entanglement, the relative entropy of entanglement and the
robustness of entanglement (from right to left). In the mixed
state case, they include some quantification of how mixed the
state is. This makes sense in the problem of state discrimination,
since the more mixed the states are, the fewer orthogonal states
there can be for a given Hilbert space dimension $D$. We will
later show that the quantities in eq.(\ref{eqn: Theorem 2}) are
equivalent for GHZ sates (these are multipartite states defined
originally in \cite{GHZ}).

To prove the relationship to the robustness of entanglement we
must first write $d(\rho)$ in a more convenient form. We can
rewrite condition $(i)$ in (\ref{eqn: DEF d1}), as $\langle
\psi|\omega|\psi\rangle \leq {\rm tr}\{\rho\omega\} ~ \forall ~
|\psi\rangle$. By considering the spectral decomposition of
$\omega$, it follows that $\omega$ can always be rewritten in the
form $\omega = \lambda |P| \frac{P}{|P|}+(1-\lambda |P|)\Delta$
with the additional conditions ${\rm tr}\{P\Delta\}=0$ and
$\lambda \geq \langle \psi|\omega|\psi\rangle~ \forall~
|\psi\rangle$. We can then rewrite
\begin{eqnarray}
\label{eqn: DEF d2}
    &d(\rho)= \min (1/\lambda) \nonumber \\
    &{\rm such ~that~} \exists {\rm ~a ~state~ }\Delta {\rm ,~satisfying} \nonumber \\
    &\omega = \lambda |P|
\frac{P}{|P|}+(1-\lambda |P|)\Delta \in SEP, \nonumber\\
    &{\rm tr}\{P\Delta\}=0, \lambda \geq\langle \psi|\omega|\psi\rangle
\forall |\psi\rangle.\nonumber \\
\end{eqnarray}
We can now compare this to the {\it global robustness of
entanglement} $R_g(\rho)$ \cite{Vidal99}.
\begin{eqnarray}
\label{eqn: DEF robustness}
    &R_g(\rho):= \min t \nonumber \\
    &{\rm such ~that~} \exists {\rm ~a ~state~ }\Delta {\rm ,~satisfying} \nonumber \\
    &{1 \over 1+t }(\rho + t\Delta)\in SEP.
\end{eqnarray}
We can understand this as the minimum (arbitrary) noise $\Delta$
that we need to add to make the state separable.

We can see that the global robustness of entanglement of the
support of state $\rho$, $R_G(P/|P|)$, is very similar in
definition to $d(\rho)$ above, (\ref{eqn: DEF d2}). The crucial
difference being the removal of the two conditions in the last
line of (\ref{eqn: DEF d2}). Since relaxing conditions can only
lead to a lower minimum, we can see that
\begin{equation}
d(\rho) \geq r(\rho) := |P|\left[1 + R_g(P/ |P|)\right], \label{eqn:
bound Robustness}
\end{equation}
proving the left inequality of theorem 2.

For the centre and right inequalities of theorem 2, we consider the
two quantities separately.

The {\it relative entropy of entanglement} is defined as
\cite{Vedral98},
\begin{eqnarray}
\label{eqn: DEF rel entropy of entanglement}
    E_R(\rho):=\min_{\omega\in SEP} S(\rho||\omega),
\end{eqnarray}
where $S(\rho||\omega) = -S(\rho) - {\rm tr}\{\rho \log_2\omega\}$
is the relative entropy and $S(\rho)$ is the von Neumann entropy.
From the definition of $R_g(\rho)$, we know that for some state
$\Delta$, the state given by: $\omega_i := [P_i/|P_i| +
R_g(P_i/|P_i|) \Delta ] /[ 1 + R_g(P_i/|P_i|)]$ is a separable
state. Hence the following inequalities must hold:
\begin{eqnarray}
E_R(\rho_i) &+& S(\rho_i) \nonumber\\
&\leq& - {\rm tr}\left\{\rho_i \log_2\left({ P_i/|P_i| +
R_g(P_i/|P_i|)\Delta \over 1 + R_g(P_i/|P_i|)}
\right)\right\} \nonumber \\
 &\leq& - {\rm tr}\left\{\rho_i
\log_2\left({ P_i/|P_i| \over 1 + R_g(P_i/|P_i|)} \right)\right\}\
\nonumber \\
&=& \log_2\left[|P_i|\left(1+R_g\left(P_i /
|P_i|\right)\right)\right],
\end{eqnarray}
where the third line follows from the monotonicity of the
logarithm, which states that $\log (A + B) \geq \log(A)$ whenever
$B \geq 0$, for two operators $A,B$ \cite{Bhatia}. The last line
is true even if $\rho_i$ is any state in the span of $P_i$. Hence:
\begin{eqnarray}
2^{E_R(\rho_i) + S(\rho_i)} \leq r(\rho_i). \label{eqn: bound ES}
\end{eqnarray}

We call the {\it geometric measure} $G(\rho)$
\begin{equation}
\label{geometric entanglement}
    G(\rho):=- \log_2 \left\{ \max_{\omega \in{SEP}}
    {\rm tr}\{\rho\omega\}\right\}.
\end{equation}
In the case of pure states, this reduces to the geometric measure
of entanglement \cite{Wei03}. However, for mixed states, this is
not an entanglement monotone (for example it is maximised by the
maximally mixed state). We immediately see that this would be
equivalent (up to log) to $d(\rho)$ in (\ref{eqn: DEF d1}) if we
were to drop condition $(i)$. Hence we have $d(\rho) \geq
2^{G(\rho)}$. However, it is possible to show a stronger bound. In
\cite{Wei04} it was shown that in the pure state case that
$G(\rho)$ is bounded from above by the relative entropy of
entanglement. We use the same simple concavity arguments now for
the mixed state case. By definition $E_R(\rho_i) + S(\rho_i)
=-\max_{\omega \in{SEP}} {\rm tr}\{\rho \log_2 \omega\}$. By
concavity of the logarithm, we have for all $\rho, \omega$, ${\rm
tr}\{\rho \log_2 \omega\}\leq {\rm tr}\{\rho\omega\}$. Thus
\begin{eqnarray}
E_R(\rho) +S(\rho) \geq G(\rho) \label{eqn: bound G}
\end{eqnarray}
Combining (\ref{eqn: bound Robustness}), (\ref{eqn: bound ES}) and
(\ref{eqn: bound G}) we get theorem 2. $\square$

We will now look at how we can use our necessary conditions to
bound the maximum number of states that can be discriminated
locally. Combining theorem 1 and 2, and dividing by $N$, we obtain
the following corollary.

{\bf Corollary:} The number of states $N$ that can be
discriminated perfectly by LOCC is bounded by

\begin{eqnarray} \label{eqn: bound N}
N \leq D/\overline{d(\rho_i)} \leq D/\overline{r(\rho_i)} &\leq&
D/\overline{2^{E_R(\rho_i) + S(\rho_i)}}  \leq D /
\overline{2^{G(\rho_i)}},\nonumber \\
\end{eqnarray}
where $\overline{x_i}:=1/N\sum_{i=1}^N x_i$, denotes the
`average'.

Hence, in the pure state case, where the bounding quantities
reduce to the geometric measure of entanglement, the relative
entropy of entanglement and the robustness of entanglement (from
right to left), we can interpret these three distance like
entanglement measures as bounds on the number of pure states that
we can discriminate perfectly by LOCC (see Fig \ref{Fig. 1}).

Given this hierarchy of bounds (\ref{eqn: bound N}), we can apply
known results from entanglement theory to find some bounds on $N$,
one of which we will show is tight. Firstly, the robustness of
entanglement is completely solved for pure bipartite states
\cite{Vidal99}. For a state with Schmidt decomposition
$|\psi\rangle=\sum_i \alpha_i |ii\rangle$, the robustness was
found to be $R_g(\psi)= (\sum_i \alpha_i)^2-1$. We can immediately
put this into (\ref{eqn: bound N}). For instance, if we have a set
of pure bipartite states all with the same entanglement, ($Bi$),
we have
\begin{eqnarray} \label{eqn: bound Bipartite}
    N(Bi) \leq d_1 d_2/(\sum_i \alpha_i)^2,
\end{eqnarray}
where $d_1, d_2$ are the dimensions of the Hilbert spaces and
$\alpha_i$ are the Schmidt coefficients for any one of the states
in the set. This has the consequence that it is impossible to
distinguish more than $d$ maximally entangled states (where $d$ is
the dimension of one subspace, then $(\sum \alpha_i)^2=d$),
reproducing a known result \cite{Ghosh0104,Nathanson04}.

In the multiparty case, we know from Wei {\it et al} \cite{Wei04}
that for the $m$-party W state $|W\rangle := |00...01\rangle
+|00..10\rangle + .. + |01..00\rangle + |10..00\rangle$ and GHZ
state $|GHZ\rangle:= |0\rangle^{\otimes m}+|1\rangle^{\otimes m}$,
the relative entropy of entanglement and the geometric measure
coincide and are given by $E_R(|GHZ\rangle)=E_G(|GHZ\rangle)=1$
and $E_R(|W\rangle)=E_G(|W\rangle)=\log_2(m/(m-1))^{(m-1)}$.
Therefore, for any set of states where the state with the average
geometric measure (or the lowest) is that of GHZ or W, we have
\begin{eqnarray}
\label{eqn: bound GHZW}
    N(GHZ) &\leq& 2^{m-1} \nonumber\\
    N(W)   &\leq& 2^m \left( (m-1)/m \right) ^{(m-1)}.
\end{eqnarray}

In fact, if we now call $N(\mathcal{S}_{GHZ})$ the {\it{maximum}}
number of states, in a set made of {\it all} GHZ type states (i.e.
GHZ up to local unitary transformations), that can be
discriminated perfectly by LOCC, then we can show
$N(\mathcal{S}_{GHZ}) = 2^{m-1}$ by explicit construction. We form
a set of states $\mathcal{S}_{GHZ}=\{|GHZ_i\rangle:=\one \otimes
U_i|GHZ\rangle\}_{i=1}^{2^{m-1}}$ by local unitaries $\{U_i\}$
over $m-1$ parties. The $\{U_i\}$ are formed from all the possible
combinations of products of the identity and $\sigma_x$ Pauli
operations, e.g. $U_1=\one^{\otimes m-2}\otimes \sigma_x$, giving
a set of $\sum_{k=0}^{m-1}{m-1\choose k}=2^{m-1}$ states. It is
easy to check that these can be discriminated by making local
$\sigma_z$ measurements. Calling $\mathcal{S}_W$ a set of states
equal to the W state up to local unitary transformations, with
(\ref{eqn: bound GHZW}) gives
\begin{eqnarray}
\label{eqn: ineq GHZW} N(\mathcal{S}_{W}) < N(\mathcal{S}_{GHZ}).
\end{eqnarray}

We also note that if we can find such a bound by any of the
entanglement measures in (\ref{eqn: bound N}) and show it is
tight, those measures below it in the hierarchy are equal. The GHZ
case is such an example giving $R_G(|GHZ\rangle)=1$, and is one of
the few cases where the global robustness of entanglement is known
for multiparty systems. We round off the examples by showing
another simple known result. If even one state in a complete basis
is entangled, then (\ref{eqn: bound N}) shows that the basis
cannot be discriminated perfectly \cite{Horodecki03}.

The simplicity of the basis for the proofs of main results here
allows it to be used with other necessary conditions on LOCC
measurements. The condition of separability (\ref{eqn: condition
separable}), for example, may be changed to more tractable
conditions such as positivity of partial transpose or
bi-separability \cite{Rains99}. It can easily be seen that these
conditions would lead to analogous bounds to those derived above.
In the case of bi-separability, the example of bipartite states
above shows that, for pure states, it is always possible to give
some easily computable bound.

We have given an interpretation of the global robustness of
entanglement, the relative entropy of entanglement and the
geometric measure of entanglement as bounds on the number of pure
states that can be discriminated perfectly by LOCC. Our general
mixed state results imply that the presence of entanglement
guarantees a certain minimal level for this difficulty.  The
difficulty of LOCC state discrimination is an important
consideration in various quantum information tasks, (e.g. quantum
data hiding\cite{Terhal01}), which may give more uses of these
results. This is the topic of ongoing investigations. In this
direction, it is also possible to extend theorem 1 to the case of
imperfect discrimination. This leads to bounds on the LOCC
accessible information, as in \cite{Nathanson04,DiVincenzo02},
which will be presented in a separate paper.

We thank Keiji Matsumoto and Martin Plenio for useful discussions.
This work was sponsored by the Asahi Glass Foundation, the JSPS,
Leverhulme Trust, and the Royal Commission for the Exhibition of
1851.


\begin{thebibliography}{99}
%
\bibitem{Horodecki01} M. Horodecki, Quant. Inf. Comp. {\bf 1},
3 (2001).
%
\bibitem{Dur00} W. D{\"u}r, G. Vidal and J. I. Cirac, Phys. Rev. A
{\bf 62}, 062314 (2000); S. Ishizaka and M. B. Plenio,
quantu-ph/0503025.
%
\bibitem{Vidal99} G. Vidal and R. Tarrach, Phys. Rev. A {\bf 59},
141 (1999); A. W. Harrow and M. A. Nielsen, Phys. Rev. A {\bf 68},
012308 (2003); M. Steiner, Phys. Rev. A {\bf 67}, 054305 (2003).
%
\bibitem{Vedral98} V. Vedral and M. B. Plenio, Phys. Rev. A {\bf
57}, 1619 (1998).
%
\bibitem{Wei03} A. Shimony, Ann. NY. Acad. Sci {\bf 755}, 675 (1995).
; H. Barnum and N. Linden, J. Phys. A: Math. Gen. {\bf 34}, 6787
(2001); T-C. Wei and P. M. Goldbart, Phys. Rev. A {\bf 68}, 042307
(2003).
%
\bibitem{Bennett84Ekert91} C. H. Bennett and G.
Brassard, in Proc. IEEE International Conference on Computers,
Systems and Signal Processing (IEEE Press, New York,1984); A. K.
Ekert, Phys. Rev. Lett. {\bf 67}, 661 (1991).
%
\bibitem{Hayden04Watrous05} P. Hayden and C. King, quant-ph/0409026; J. Watrous, quant-ph/0503092.
%
\bibitem{Cirac99} J. I. Cirac, A. K. Ekert, S. F. Huelga and C.
Macchiavello, Phys. Rev. A {\bf 59}, 4249 (1999).
%
\bibitem{Terhal01} B. M. Terhal, D. P. DiVincenzo and D. W. Leung, Phys. Rev. Lett. {\bf 86}, 5807
(2001).
%
\bibitem{Bennett99} C.H. Bennett, D.P. DiVincenzo, C.A. Fuchs, T. Mor,
E. Rains, P.W. Shor, J.A. Smolin and W.K. Wootters, Phys. Rev. A.
{\bf 59}, 1070 (1999).
%
\bibitem{Walgate00Virmani01} J. Walgate, A. J. Short, L. Hardy and V. Vedral,
Phys. Rev. Lett. {\bf  85}, 4972 (2000); S. Virmani, M. F. Sacchi,
M. B. Plenio, D. Markham, Phys. Lett. A {\bf 288}, 62 (2001); Z.
Ji, H. Cao and M. Ying, Phys. Rev. A {\bf 71}, 032323 (2005).
%
\bibitem{Chen02-Hillery02} M. Hillery and J. Mimih, quant-ph/0210179; P.X. Chen and
C.Z. Li, Phys. Rev. A {\bf 68},062107 (2003); S. de Rinaldis,
quant-ph/0304027; S. Virmani and M. B. Plenio, Phys. Rev. A {\bf
67}, 062308 (2003); F. Anselmi, A. Chefles and M. B. Plenio, New
J. Phys. {\bf 6}, 164 (2004); A. Chefles, Phys. Rev. A {\bf 69},
050307(R) (2004); H. Fan, Phys. Rev. Lett. {\bf 92}, 177905
(2004).
%
\bibitem{Badziag03etal}P. Badzi\c{a}g, M. Horodecki, A. Sen, U. Sen, Phys. Rev. Lett. {\bf 91}, 117901
(2003); M. Horodecki, J. Oppenheim, A. Sen(De) and U. Sen, Phys.
Rev. Lett. {\bf 93}, 170503 (2004); S. Ghosh, P. Joag, G. Kar, S.
Kunkri and A. Roy, quant-ph/0403134 (2004).
%
\bibitem{NoteSupport} The support of a state $\rho$, with eigen-decomposition $\rho = \sum_i \alpha_i
|i\rangle\langle i|$ is given by $P=  \sum_i |i\rangle\langle i|$.
%
\bibitem{GHZ}D. M. Greenberger, M. Horne, and A. Zeilinger, {\it Bell's Theorem,
Quantum Theory, and Conceptions of the Universe}, edited by M.
Kafatos (Kluwer, Dordrecht, 1989), p. 69.

%
\bibitem{Bhatia} R. Bhatia, Matrix Analysis, Springer Graduate Texts in Mathematics vol.
169 (1991).
%
\bibitem{Wei04} T-C. Wei, M. Ericsson, P. M. Goldbart and W. J. Munro, Quant. Inform. Compu. {\bf 4}, 252 (2004).
%
\bibitem{Ghosh0104} S. Ghosh, G. Kar, A. Roy, A. Sen (De) and U.
Sen,Phys. Rev. Lett {\bf87}, 277902 (2001); S. Ghosh, G. Kar, A.
Roy and D. Sarkar, Phys. Rev. A {\bf 70}, 022304 (2004).
%
\bibitem{Nathanson04} M. Nathanson, quant-ph/041110.
%
\bibitem{Horodecki03}M. Horodecki, A. Sen (De), U. Sen and K. Horodecki, Phys. Rev. Lett. {\bf 90}, 047902
(2003).
%
\bibitem{Rains99} E. M. Rains, Phys. Rev. A {\bf 60}, 179 (1999);
Phys. Rev. A {\bf 63}, 019902(E) (2000).
%
\bibitem{DiVincenzo02} D.P. DiVincenzo, D. W. Leung and B. M. Terhal,
IEEE Trans. Inform. Theory {\bf 48}, 580 (2002).
%
\end{thebibliography}

\end{document}